# Turn-of-the Year Affect in Gold Prices: Decomposition Analysis[1]

**Asst. Prof. Osman GÜLSEVEN**
American University of Middle East
osman.gulseven@aum.edu.kw

***Abstract***

In this article, we examine whether the gold market returns is subject to abnormally positive or negative returns in some months of the calendar year. We derive the percentage monthly returns of gold prices denoted in both major global currencies and the currencies of the largest gold consumer markets. The statistical analysis and the decomposition techniques suggests gold prices show some seasonal behaviour during the turn of the year. Our findings for these months are robust to the chosen currency, albeit with some differences in monthly returns between highly correlated major currencies and loosely correlated gold demanding currencies. We observe a strong cyclical behaviour in gold markets during the turn-of-the-year period. January is likely to offer the highest return whereas significant negative returns are expected in July.

## INTRODUCTION

Gold is one of the most commonly held commodities. People invest in gold for different reasons. Most common reasons why people invest in gold is due its role store of value, portfolio diversifier, and also as hedge role against financial shocks. These roles undertaken by gold do not have strictly defined limitations. Gold can act as a store of value in a diversified portfolio which is hedged against financial shocks (Lu and Chan, 2014). The role of gold in emerging market currencies is more prominent as gold also offers exposure to stabilizing currency fluctuations. Thus, in case of a local currency devaluation during a financial turmoil, the value of gold in local currency is likely to increase thereby effectively acting as both a hedge and safe heaven (Gulseven and Ekici, 2016; Miyazaki and Hamori, 2013). However, Pukthuanthong and Roll (2011) suggest that for strong currencies such as Euro, the price of gold in Euro moves in line with the value of Euro, which magnifies the effect of currency depreciations.

In recent years, the financial turmoil in global markets has attracted a large number of market participants in the gold market. The securitization of gold markets contributed to this increased public interest in gold. As such, there are millions of buyers and sellers acting simultaneously to news in the gold market. In terms of hedging role, the gold-index exchange trade funds (ETFs) share the same properties as the physical gold (Pullen et al., 2014). Similar to other large commodity markets, gold has a well-established market where any abnormal returns are expected to disappear over time (Charles, et al., 2015). However, the dynamics of the gold market substantially differs from that of other commodities, such as oil, as gold can be easily stored. Gold price is also highly affected by speculations and herd behaviour in the financial market.

Nevertheless, there is a strong tendency for the gold prices to increase over time. At the peak of Euro crises in 2012, gold was trading for as much as 1750 USD per ounce. The prices have since dropped back. Since 1978, the price of gold has increase by 4-fold from 226 USD per ounce to 1234 USD per ounce. The prices have experienced more or less similar paths when denoted in terms of global currencies European Euro (EUR), Japanese Yen (JPY), Great Britain Pound (GBP), Canadian Dollar (CAD) and Swiss Franc (CHF). In consumer countries, there was a bit variety in gold prices. The gold prices in tightly controlled currencies such as Chinese renminbi (CNY), Saudi Riyal (SAR), UAE Dirham (AED), have followed a similar track with global currencies. However, the gold prices in weaker consumer currencies such as Indian Rupee (INR), Turkish Lira (TRY), and Indonesian Rupiah (IDR) have increased by several folds due to local currency devaluation. Thus, gold acted as one of the best instruments against currency devaluation in countries with weak currencies. Interestingly, some also consider gold as a hedge against USD depreciation (Capie, et al., 2005).

[1] This paper was presented as an abstract in August, 24-26 2016 at International Congress on Political Economic and Social Studies, Istanbul/Turkey.

With the introduction of exchange traded funds, gold has become more of a financial asset rather than a commodity in recent years. This transformation has received a significant attention from not only investors but also short-term speculators who would like to benefit from short term movements in gold prices. In this article, we hypothesize that one might observe calendar based market anomalies in a similar fashion to stock markets. Specifically, we test for monthly calendar based market anomalies. We apply a unique methodology based on decomposition method by decomposing our data into monthly seasonal components and a linear trend over time. Our objective is to see whether some months of the year offer consistently higher or lower returns compared to the rest of the months.

The organization of the paper is as follows. In the next section, we explain the relevant literature with a particular emphasis of seasonal analysis applications. In the data section, we explain our data, the variables used in estimation and their statistical properties. In the method section, we explain how our decomposition model is applied to the empirical gold price data denoted in 6 global major and 6 gold consumer currencies. Finally, the last section concludes with a summary of our findings and policy recommendations.

**1. Literature Review**

The research on different aspects of the gold market is pretty large and growing along with market interest. Here, we refer to here only the relevant research that is most related to our subject. See O'Connor et al. (2015) for an extensive review on many aspects of the gold market.

While there is a positive tendency in gold prices, the path of market direction is unknown. Some consider gold market as a random walk with positive random drifts over time whereas some researchers suggest gold prices can be predicted using time series analysis (Basu and Claouse, 1993). Another direction of research is to estimate the price behaviour based on other somewhat relevant market indicators such as US inflation, US dollar exchange rate, gross domestic product (GDP), etc. (Kaufmann and Winters, 1989; Cai et al., 2011). Inflation, measured by consumer price index (CPI) is thought to be the one of the most prominent factors in determining gold prices (Sharma, 2016). The effect of interest rates on gold prices are also widely studied. Gold is also thought to be cointegrated with other commodities such as oil where energy markets interact with gold markets (Kiohos and Sariannidis, 2010; Zhang, et al. 2010; Kristoufek, 2010). Narayan et al. (2010) find strong correlation between gold and oil futures. The correlation between gold and silver is also thought be strong (Batten et al., 2013).

Regional studies of gold prices in different currencies have also been of interest to several researchers. El Hedi et al. (2015) state that gold not only acts as a diversifier, but it has been a safe haven for Chinese stock market investors. However, Gencer and Musoglu (2014) state that although gold is a portfolio diversifier, the role of gold as a safe haven in Turkish markets during turmoil periods is unfounded. Kiran (2010) states that the gold returns in Turkish lira follows a covariance stationary and mean reverting process. Munir and Kok (2014) find a similar result for the Malaysian gold bullion coin prices. Market parameters in India, the largest consumer market for gold, also thought be important price determinants. Several researchers investigated the effect of Indian-based market variables such as Indian rupee US Dollar exchange rate (Jain and Ghosh, 2013); and gold prices in India (Mali, 2014). A similar research by Shahbaz et al. (2014) on Pakistani market suggests gold can be a good inflation hedge both in short and long run for Pakistanis. European gold markets are also investigated by researchers. Gallais-Hamonno, et al. (2015) state that inefficiencies exists even in the official gold markets in Paris. On the contrary, Ho (1985) finds that London gold market is incrementally efficient with respect to exchange rates.

The existence and the cause of inefficiencies as well as the predictability of the gold market is subject to serious debate. Pierdzioch et al (2014) apply real-time forecasting approach to prove that gold markets are informationally efficient. Tschoegl (1980) claim that inefficiencies even if they exist cannot be exploited for abnormal returns. On the contrary, Mehrara et al. (2014) claim that gold market is not efficient in Fama sense. This finding is partially supported by Ntim et al. (2015) who prove that while established markets show efficiency, developing markets are still price inefficient.

Xian et al. (2016) also support existence of market inefficiencies by suggesting that empirical model decomposition can serve as an effective method to analyze gold prices.

There is a large literature on the seasonal behaviour of the stock markets which can be associated with the seasonal behaviour in gold markets. Day of the week effect is one of the most investigated phenomena in the market. Berument and Kiymaz (2001) find that returns and volatilies are significantly different in different days of the week. Balaban et al. (2001) observe that there are is a significant day of the week effect among several stock market indices albeit in different directions for different countries. Recently, monthly seasonal behaviour of stock markets have also gained attention (Gulseven, 2014). Specifically researchers have concentrated on whether there exist a positive January effect (Mehdian and Perry, 2012) or a negative May effect (see Banuman and Jacobsen, 2010; Dzhabarov and Ziemba, 2010).

As stock markets and commodity markets are strongly related one would expect to see similar amount of research in commodity prices. However, although seasonal affects are widely investigated in stock markets, there are only a few studies addressing the seasonal behaviour of gold prices. Lucey (2010) claim that lunar calendar based anomalies exist in silver and gold markets. Blose and Gondhalekar (2013) find out a significantly negative skewness over the weekend period for the gold returns. Baur (2013) checks to see whether the gold market is seasonal using dummy variable based integration. The author finds out that gold returns show strong autumn seasonality where abnormal returns are observed in September and November. This positive abnormal return is supported by the Chinese data (Min and Wang, 2013). The authors state that the gold return is higher in months of February, September, and November, which is probably due to the long public holidays coinciding with these periods. Our research differs from the previous literature as we use the longest available data in 12 different currencies available dating back to 1970, applied to a multiplicative decomposition model.

## 2. Data, Methodology and Analysis

### 2.1. Data

We collected the primary data for our empirical research from the World Gold Council Global Gold Prices Database. This database offers information on end of the month gold prices in several different currencies. For our research, first we selected the top global market currencies where gold prices are commonly denoted. The currencies listed as the global major currencies include Unites States Dollar (US), European Euro (EUR), Japanese Yen (JPY), Great Britain Pound (GBP), Canadian Dollar (CAD) and Swiss Franc (CHF). The data for major currencies goes back to January 1979. We use the data from this period up to February 2016. Thus, we have 447 price and 446 return gold price observations for major currencies.

We also included the prices in terms of major gold consuming countries. The consumer currencies used in our research include Indian Rupee (INR), Chinese Renminbi (CNY), Turkish Lira (TRY), Saudi Arabian Riyal (SAR), Indonesian Rupiah (IDR), and United Arab Emirates Dirham (AED). The data for consumer currencies goes back to January 1985. Therefore, we have 374 observations for the gold prices in these currencies. Among these currencies SAR and AED are anchored to USD. Hence, we expect to see similar results for these currencies. Chinese Renminbi is subject to strict discretionary state control. INR, CNY, and TRY belong to countries that experienced significant inflation in the past decades.

The percentage arithmetic returns are calculated as following:

$$Return_t = \frac{Price_t - Price_{t-1}}{Price_{t-1}} \qquad (1)$$

### 2.2. Summary Statistics

Table 1 below presents for the relevant period the average monthly gold returns denoted in terms of major currencies listed as USD, EUR, JPY, GBP, CAD, and CHF.

**Table 1: Average monthly gold returns (%) – Major currencies**

| Month | USD | EUR | JPY | GBP | CAD | CHF |
|---|---|---|---|---|---|---|
| **1** | 1.65% | 3.39%* | 2.09% | 2.73%* | 2.24% | 3.07%* |
| **2** | -0.05% | -0.01% | 0.18% | 0.52% | -0.17% | 0.21% |
| **3** | -0.95% | -0.56% | -0.69% | -1.11% | -1.09% | -0.93% |
| **4** | 0.60% | 0.28% | 0.09% | -0.39% | -0.25% | 0.26% |
| **5** | 0.30% | 1.21% | 0.48% | 0.76% | 0.45% | 0.83% |
| **6** | -0.29% | -0.23% | -0.32% | -0.32% | -0.31% | -0.73% |
| **7** | 0.31% | -0.01% | 0.23% | -0.21% | 0.40% | -0.22% |
| **8** | 1.51% | 1.72% | 1.03% | 2.01%* | 1.70% | 1.51% |
| **9** | 2.54%* | 1.83% | 1.79% | 2.74%* | 2.17%* | 1.28% |
| **10** | -0.94% | -0.88% | -1.24% | -1.06% | -0.81% | -0.82% |
| **11** | 1.67%* | 1.70%* | 1.55% | 1.94%* | 1.99%* | 1.46% |
| **12** | -0.04% | -1.08% | -0.26% | -0.25% | 0.26% | -1.10% |
| **Average** | 0.53%* | 0.62%* | 0.41% | 0.62%* | 0.55%* | 0.41% |

*Statistically Significant Returns at 95%

Based on the summary statistics, one can claim that for all major currencies except JPY and CHF, the average monthly gold returns are significantly positive. On a monthly basis, January is a positive month for gold denoted in terms of EUR, GBP, and CHF. August is a positive month for gold denoted in terms of GBP. September is a positive month for gold denoted in terms of USD, GBP, and CAD. November is a positive month for gold denoted in terms of USD, EUR, GBP, and CAD. March, June, and October tend to negative months for all currencies although not significant in statistical terms.

Table 2 below defines the correlation matrix on monthly nominal gold returns defined in terms of major currencies. The left side of the graph shows the correlation matrix in terms of nominal gold prices and the right side shows the correlation matrix in terms of percentage gold returns.

**Table 2: Correlation matrix on gold prices and returns – Major currencies**

| **Gold Price Correlation** | | | | | |
|---|---|---|---|---|---|
| | **USD** | **EUR** | **JPY** | **GBP** | **CAD** |
| **EUR** | 0.98 | | | | |
| **JPY** | 0.81 | 0.81 | | | |
| **GBP** | 0.99 | 0.99 | 0.77 | | |
| **CAD** | 0.99 | 0.98 | 0.80 | 0.98 | |
| **CHF** | 0.92 | 0.92 | 0.93 | 0.91 | 0.90 |
| **Gold Return Correlation** | | | | | |
| | **USD** | **EUR** | **JPY** | **GBP** | **CAD** |
| **EUR** | 0.84 | | | | |
| **JPY** | 0.83 | 0.81 | | | |
| **GBP** | 0.86 | 0.90 | 0.79 | | |
| **CAD** | 0.93 | 0.84 | 0.77 | 0.84 | |
| **CHF** | 0.80 | 0.94 | 0.82 | 0.86 | 0.78 |

*All correlations are statistically significant at 95%

All correlations are significant and there seems to be almost perfect correlation between nominal prices. This is particularly evident in USD, EUR, GDP, and CAD. However, when we measure the correlation in terms percentage returns the correlations become weaker. Here, the proximity of the currencies become a dominant factor in correlation strength. Thus, gold prices in CHF is highly correlated with gold prices in EUR, which itself is highly correlated with GBP. Prices in CAD and USD are also strongly correlated.

The prices in terms of gold demanding countries follow a somewhat similar trend. However, the average monthly returns are significantly higher in currencies such as TRY and IDR which experienced persistent depreciation in the recent decades.

**Table 3: Average monthly gold returns (%) – Demand currencies**

| Month | INR | CNY | TRY | SAR | IDR | AED |
|---|---|---|---|---|---|---|
| **1** | 1.67%* | 3.11%* | 3.77%* | 1.27% | 5.05%* | 1.26% |
| **2** | 0.84% | 0.50% | 4.76%* | 0.53% | 0.10% | 0.49% |
| **3** | 0.05% | -0.33% | 3.62%* | -0.15% | -0.16% | -0.26% |
| **4** | 0.46% | 0.16% | 2.98%* | 0.30% | -0.07% | 0.46% |
| **5** | 1.33%* | 0.45% | 2.68%* | 0.10% | 1.85% | 0.07% |
| **6** | 0.07% | -0.82% | 1.47% | -0.29% | 0.39% | -0.39% |
| **7** | 1.25% | 0.91% | 1.52% | 0.44% | 0.06% | 0.48% |
| **8** | 1.82% | 0.97% | 3.36%* | 0.81% | 1.68% | 0.79% |
| **9** | 2.35%* | 1.84% | 4.16%* | 2.21%* | 4.95%* | 2.18%* |
| **10** | -0.40% | -0.56% | 1.57%* | -0.73% | -1.29% | -0.75% |
| **11** | 2.4%* | 2.16%* | 3.32%* | 1.52%* | 2.64%* | 1.52%* |
| **12** | -0.47% | 0.44% | 1.93% | -0.22% | 1.18% | -0.22% |
| **Average** | 0.94%* | 0.74%* | 2.94%* | 0.49%* | 1.37%* | 0.47%* |

On average all gold prices in all demand currencies are significantly positive. Specifically for Turkey even individual monthly returns are significantly positive which is probably due to the hyperinflation experienced in this country during the last decades. On a monthly basis, January is significantly positive for gold prices denoted in INR, CNY, TRY, and IDR. September is positive for all demand currencies except CNY and November is significantly positive for gold in all demand currencies.

**Table 4: Correlation matrix on gold prices and returns – Demand currencies**

| **Gold Price Correlation** | | | | | | |
|---|---|---|---|---|---|---|
| | **USD** | **INR** | **CNY** | **TRY** | **SAR** | **IDR** |
| **INR** | 0.97 | | | | | |
| **CNY** | 0.97 | 0.96 | | | | |
| **TRY** | 0.95 | 0.99 | 0.94 | | | |
| **SAR** | 0.99 | 0.97 | 0.97 | 0.95 | | |
| **IDR** | 0.95 | 0.99 | 0.95 | 0.99 | 0.95 | |
| **AED** | 0.99 | 0.97 | 0.97 | 0.95 | 0.99 | 0.95 |
| **Gold Return Correlation** | | | | | | |
| | **USD** | **INR** | **CNY** | **TRY** | **SAR** | **IDR** |
| **INR** | 0.87 | | | | | |
| **CNY** | 0.80 | 0.70 | | | | |
| **TRY** | 0.59 | 0.60 | 0.55 | | | |
| **SAR** | 0.99 | 0.86 | 0.50 | 0.59 | | |
| **IDR** | 0.51 | 0.49 | 0.42 | 0.41 | 0.52 | |
| **AED** | 0.99 | 0.86 | 0.80 | 0.59 | 1 | 0.52 |

*All correlations are statistically significant at 95%

An interesting relationship is observed in gold prices and returns when prices are denoted in demand currencies. Compared to prices listed in major global currencies, the correlations among nominal prices are higher whereas the correlations in price returns are much lower. Only SAR and AED resemble a close price-return behaviour to that of USD as those currencies are anchored to USD. However, the return relationships in gold prices measured as correlations between USD, INR, CNY, and IDR are much lower. Thus, the correlation in monthly gold price returns denoted in these currencies are significantly lower although they all refer to the monthly return in gold investment. This phenomena suggest that these countries have their own unique macroeconomic dynamics affecting their currencies in a diversified way. Nevertheless, one thing common in all markets is that the gold prices have a significant tendency to go up in local currencies.

### 2.3. Model

While the descriptive statistical analysis of data suggests some sort of abnormal returns during specific months, these returns are based on simple statistical analysis without reference to any time series analysis. However the price data over time is autocorrelated due to the trending nature of data. Therefore, one needs to apply an analysis which allows for accommodation of autocorrelation due to trend. One such method is decomposition analysis. The decomposition analysis decomposes the data into different segments such as trend, seasonality, cycle, and irregular components. It is a practical approach with wide applications in macroeconomic time series analysis (Gooijer and Hyndman, 2006).

The trend component captures the trends in the data where the trend equation specifies whether it is positive or negative over time. The seasonal component identifies the seasonal effects in the data if they exist. Since our data is based on calendar month, we defined the seasonal length as 12, implying that if seasonal effects exist then they shall repeat themselves after 12 months. We did not impose any cyclical behaviour in the data, thereby the segment not explain by the trend and seasonal behaviour is considered as irregular component.

In the literature there are two distinct ways to decompose the time series. The first approach defines seasonal and trend components in additive form:

$$Index_t = Trend_t + Season_t + Irregular_t \qquad (2)$$

The above approach assumes an additive trend and seasonal component which is not the case with our data as we have an exponentially growing prices in the gold market. The second method defines the index variable as a multiplication form of the trend, seasonality and irregular components:

$$Index_t = Trend_t \, X \, Season_t \, X \, Irregular_t \qquad (3)$$

The above technique is called multiplicative decomposition technique, where the functional form better accommodates exponentially growing variables such as gold prices.

### 2.4. Results

In this section we present our results. We simultaneously estimate the trend and seasonality equations for each currency price. To see whether there are any differences or similarities we analyzed the results separately for major currency prices and consumer currency based prices.

#### 2.4.1 Monthly Price Decomposition - Major Currencies

As expected, the trend equations for all major currencies suggest a positive trend in gold prices. Table 5 below shows the trend equations and monthly decomposition analysis results for USD, EUR, JPY, GBP, CAD, and CHF.

The trend equation analysis suggests somewhat similar trend equations for major currencies except for JPY which has a distinctively different value compared to other currencies.

**Table 5. Monthly decomposition analysis of gold prices – Major currencies**

| Month | USD | EUR | JPY | GBP | CAD | CHF | |
|---|---|---|---|---|---|---|---|
| **1** | 1.0019 | 1.0113 | 1.0028 | 1.0028 | 1.0113 | 1.0063 | + |
| **2** | 1.0039 | 1.0084 | 1.0033 | 1.0159 | 1.0149 | 1.0129 | + |
| **3** | 1.0004 | 1.0011 | 1.0075 | 0.9993 | 0.9985 | 1.0108 | 0 |
| **4** | 0.9947 | 0.9930 | 0.9963 | 0.9930 | 0.9869 | 0.9956 | - |
| **5** | 0.9959 | 1.0061 | 1.0137 | 0.9994 | 0.9914 | 1.0084 | 0 |
| **6** | 0.9981 | 0.9966 | 0.9959 | 0.9937 | 0.9948 | 0.9906 | - |
| **7** | 0.9781 | 0.9835 | 0.9884 | 0.9866 | 0.9831 | 0.9893 | - |
| **8** | 0.9987 | 1.0121 | 0.9992 | 1.0020 | 0.9994 | 1.0038 | 0 |
| **9** | 1.0034 | 1.0072 | 0.9992 | 0.9995 | 1.0021 | 1.0061 | 0 |
| **10** | 1.0011 | 0.9909 | 0.9898 | 0.9925 | 0.9899 | 0.9870 | 0 |
| **11** | 1.0112 | 1.0071 | 1.0061 | 1.0167 | 1.0133 | 1.0050 | + |
| **12** | 1.0125 | 0.9828 | 0.9978 | 0.9987 | 1.0147 | 0.9842 | 0 |
| **MAPE** | 49 | 43 | 50 | 51 | 36 | 36 | |
| **MAD** | 230 | 177 | 29327 | 149 | 225 | 236 | |
| **MSD** | 82382 | 43466 | 1.17E+7 | 32662 | 71607 | 77341 | |
| **Constant** | 117.2 | 129.7 | 55671 | 54.4 | 216.4 | 513 | |
| **Slope** | 2.09xt | 1.537xt | 73.9xt | 1.348xt | 2.126xt | 1.035xt | |

The decomposition analysis suggests differentiated seasonal components for prices denoted in different currencies. However, there is an obvious positive tendency during January, February, and October. Negative tendencies are observed in April, June and July.

As can be seen in Graph 1, when denoted in major currencies there seems to be a cyclical behaviour in the gold prices. From November until March, there seems to be seasonal hike in gold prices, which turns to be seasonal negative returns during the rest of the year. June and July are significantly negative periods where the seasonal returns in July is almost 2% lower than the average returns in other months.

**USD Monthly Decomposition**

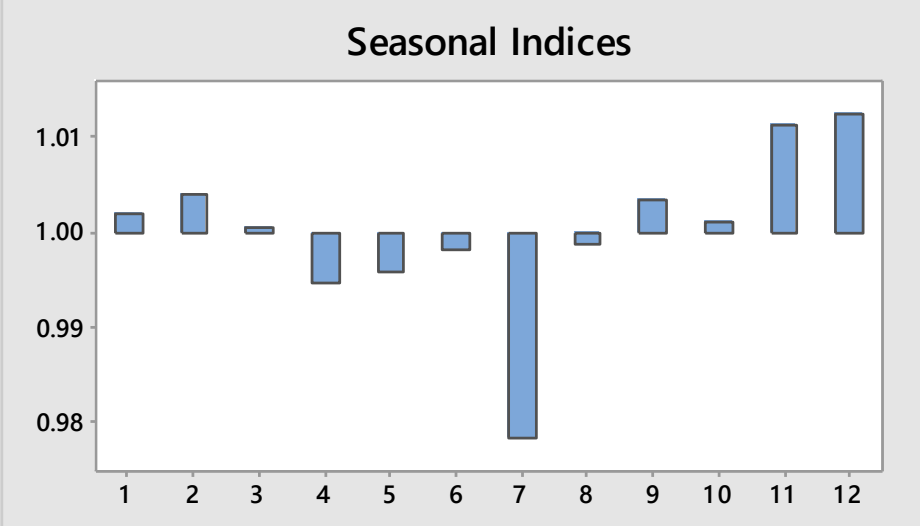

**EUR Monthly Decomposition**

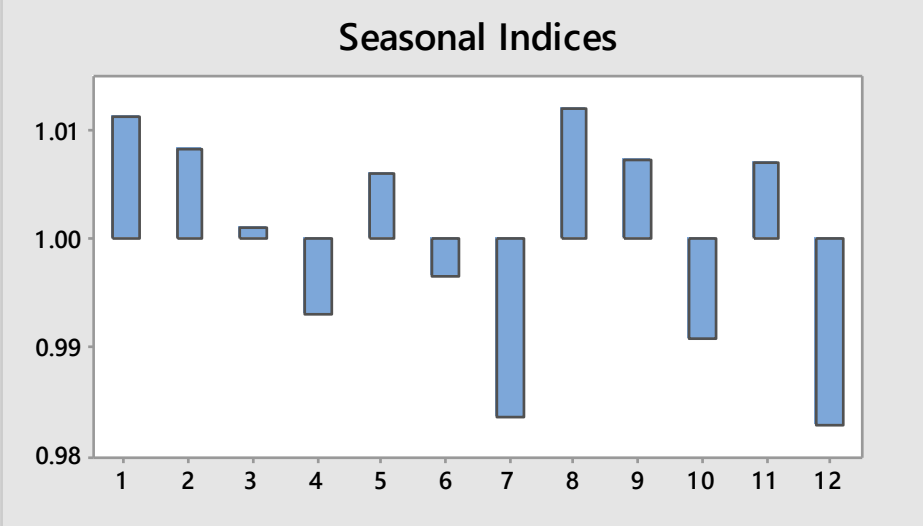

**JPY Monthly Decomposition**

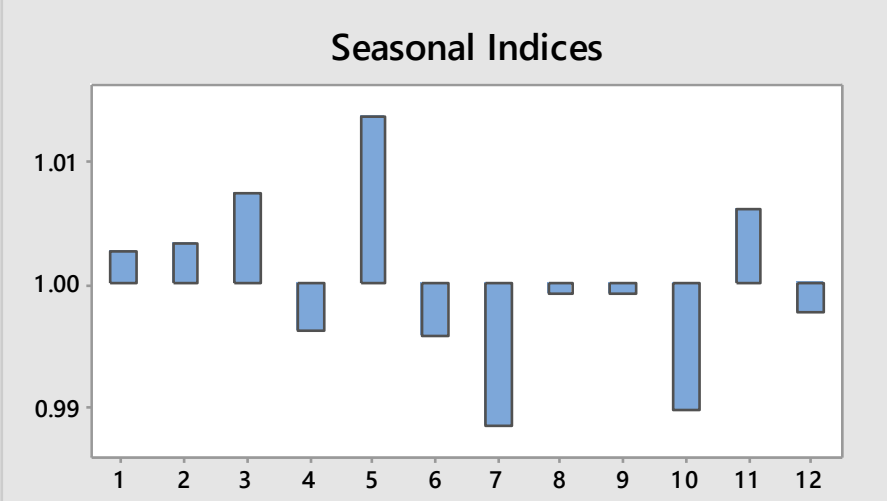

**GBP Monthly Decomposition**

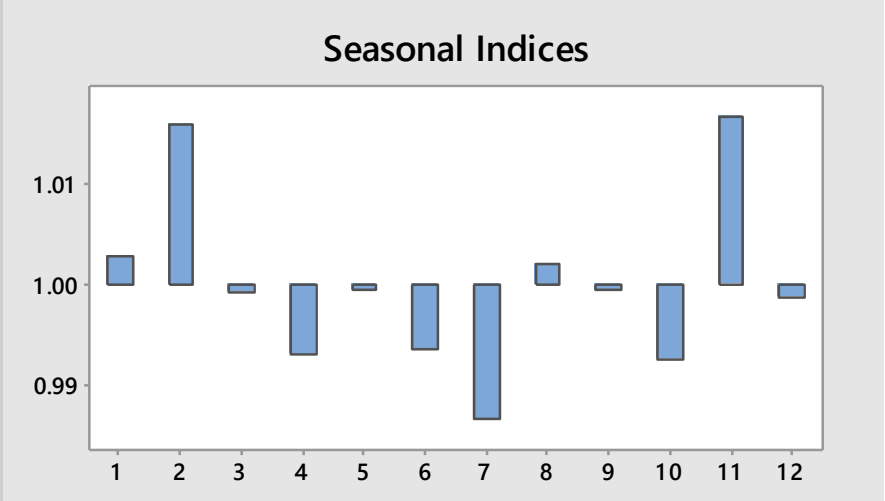

**CAD Monthly Decomposition**

**CHF Monthly Decomposition**

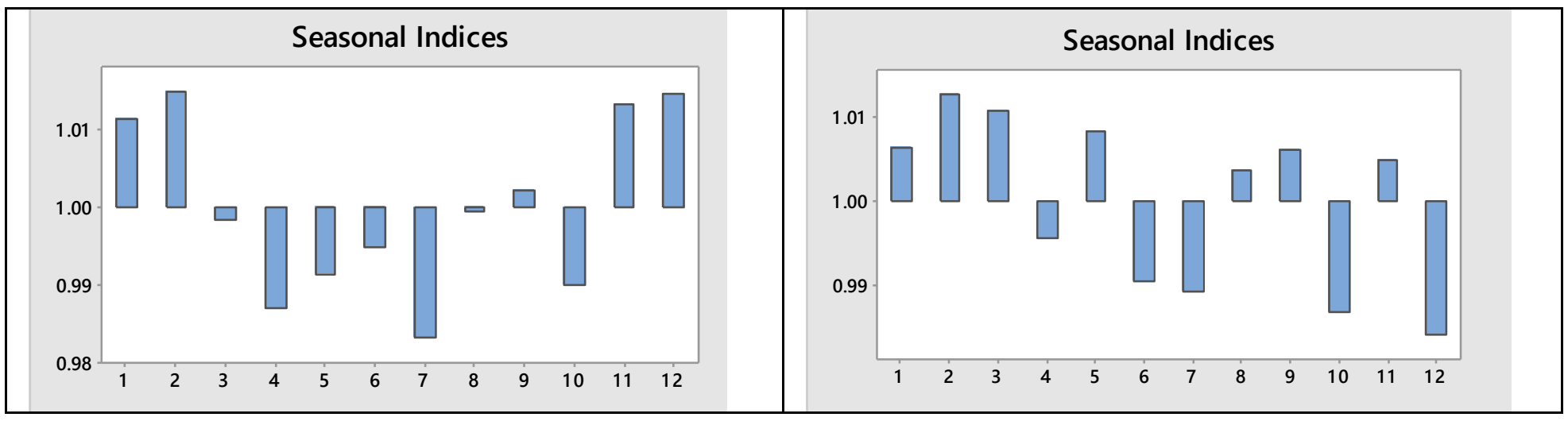

**Graph 1: Monthly decomposition analysis for percentage gold returns – Major currencies**

### 2.4.2 Monthly Price Decomposition - Demand Currencies

Similar to major currency prices, the trend equations for all demand currencies suggest a positive trend in gold prices. Table 6 below shows the trend equations and monthly decomposition analysis results for INR, CNY, TRY, SAR, IDR, and AED.

However, unlike major currencies the constants in trend equations suggest conflicting results. Only the constant prices for SAR and AED show positive value. The constant values for other currencies turned out to be negative which is probably due to the high currency depreciation experienced in these countries.

**Table 6: Monthly decomposition analysis of gold prices – Demand currencies**

| | INR | CNY | TRY | SAR | IDR | AED | |
|---|---|---|---|---|---|---|---|
| **1** | 0.9997 | 1.0108 | 1.0094 | 1.0052 | 1.0174 | 1.0047 | + |
| **2** | 1.0054 | 1.0132 | 1.0122 | 1.0089 | 1.0084 | 1.0093 | + |
| **3** | 0.9926 | 0.9998 | 1.0090 | 1.0006 | 0.9958 | 1.0007 | 0 |
| **4** | 0.9855 | 0.9924 | 1.0085 | 0.9953 | 0.9877 | 0.9952 | 0 |
| **5** | 0.9935 | 0.9992 | 1.0080 | 0.9972 | 0.9975 | 0.9974 | - |
| **6** | 0.9992 | 0.9982 | 1.0046 | 0.9985 | 1.0007 | 0.9982 | - |
| **7** | 0.9807 | 0.9880 | 0.9833 | 0.9834 | 0.9709 | 0.9835 | - |
| **8** | 1.0073 | 0.9960 | 1.0044 | 0.9972 | 0.9987 | 0.9972 | 0 |
| **9** | 1.0143 | 0.9993 | 0.9930 | 1.0007 | 0.9969 | 1.0007 | 0 |
| **10** | 1.0004 | 0.9981 | 0.9917 | 1.0017 | 0.9966 | 1.0020 | 0 |
| **11** | 1.0154 | 0.9987 | 0.9941 | 1.0027 | 1.0210 | 1.0027 | 0 |
| **12** | 1.0061 | 1.0066 | 0.9817 | 1.0087 | 1.0086 | 1.0086 | + |
| **MAPE** | 80 | 34 | 32790 | 48 | 4.21E+06 | 48 | |
| **MAD** | 11300 | 1059 | 467 | 828 | 1.02E+10 | 812 | |
| **MSD** | 185E+6 | 1876431 | 283600 | 50356 | 7.55E+16 | 9489 | |
| **Constant** | -12102 | -75 | -795 | 228 | -3.2E5 | 228 | |
| **Slope** | 208xt | 21.9xt | 9.38xt | 11.09xt | 43535xt | 10.85xt | |

While there is a big deal of seasonal components, one can observe a positive seasonal value for December, January, and February. Negative price sentiments are observed from May to July.

As can be seen in Graph 2, similar to major currency based prices when denoted in demand currencies there seems to be a cyclical behaviour in the gold prices albeit with slight shifts in months. From December until February, there seems to be seasonal hike in gold prices, which turns into seasonally negative returns during the rest of the year. May, June and July are significantly negative periods where the seasonal returns in July is almost 2% to 3% lower than the average returns in other months.

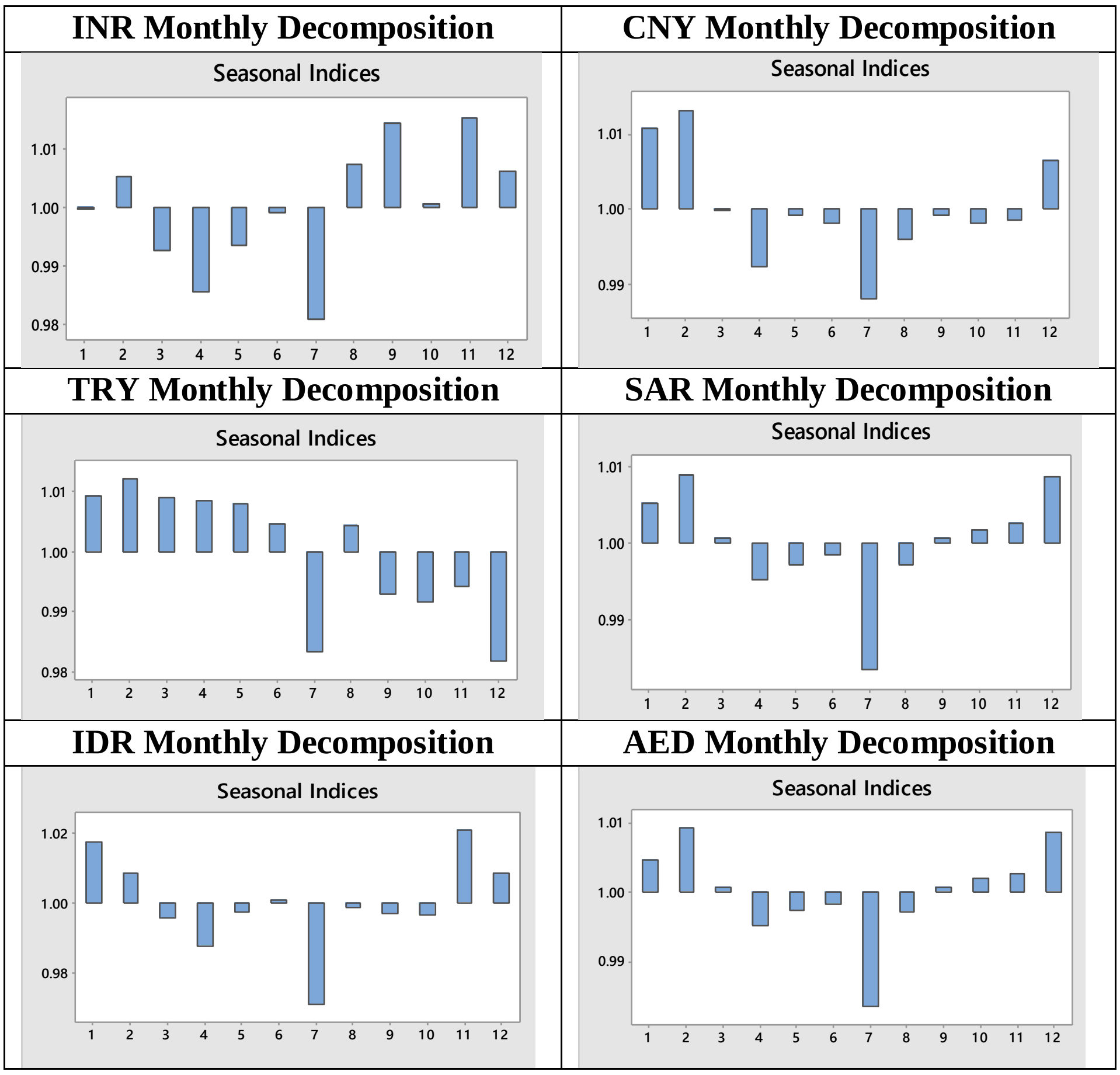

**Graph 2: Monthly decomposition analysis for gold returns (%) - Demand currencies**

## CONCLUSION

Gold has a long history as an investment instrument dating back to history of civilizations. Even after the establishments of modern financial markets the gold appetite has continued for many investors around the world.

In this article, we have tested whether the returns in gold investment is superior in some specific months of the year when measured in different currencies. Our results for gold returns when measured in major currencies suggested November, January, and February tend to be offer positive returns whereas March, June, and July tend to offer seasonally negative returns. These results are valid for gold prices measured in USD, EUR, JPY, GBP, CAD, and CHF. We also tested for abnormal gold returns when prices are measured in major demand currencies. The results for gold price returns measured in INR, CNY, TRY, SAR, IDR, and AED suggested a very similar outcome. Prices tend to be positive during the months of December, January, and February and negative during the months of May, June, and July.

The arrangement of abnormally positive months, followed by mixed months, followed by abnormally negative months, and followed by abnormally positive months suggest that gold prices indeed move in a seasonal fashion. With a few exceptions the gold returns tend to be higher during the turn-of-the year period and slightly lower during the middle of the year period. Our results provide partial support for the “autumn effect” in gold returns suggested by Baur (2013). However, using a wider range of data both in terms of duration and variables, we find strong support for the “winter effect” in gold returns. The strong winter effect is probably due to several factors coinciding at this time of year. In the Northern India November coincides with the wedding season and in the Southern India January is known as the time to buy gold for cultural reasons. This time of the year is also the Christmas period where global gold demand for ornament tends to be high. The turn-of-the-year period returns is also a known phenomenon in global stock markets. There is also a possibility that the same factors affecting the rise in stock markets also affect the global gold prices in a similar fashion.